# Integrating RDF into Hypergraph-Graph (HG(2)) Data Structure

Shiladitya Munshi
Meghnad Saha Institute of Technology
Kolkata, India
Web Intelligence and
Distributed Computing research Lab
Golfgreen, Kolkata: 700095, India
Email:shiladitya.munshi@yahoo.com

Ayan Chakraborty
Techno India College of Technology
Kolkata, India
Web Intelligence and
Distributed Computing research Lab
Golfgreen, Kolkata: 700095, India
Email: achakraborty.tict@gmail.com

Debajyoti Mukhopadhyay
Maharastra Institute of Technology
Pune, India
Web Intelligence and
Distributed Computing research Lab
Golfgreen, Kolkata: 700095, India
Email: debajyoti.mukhopadhyay@gmail.com

*Abstract*—Current paper discusses the methodologies involved in integrating Resource Description Framework (RDF) into a HyperGraph - Graph (HG(2)) data structure in order to preserve the semantics of the information contained in RDF document for dealing future cross platform information portability issues. The entire semantic web is mostly dominated by few information frameworks like RDF, Topic Map, OWL etc. Hence semantic web currently faces the problem of non existence of common information meta-model which can integrate them all for expanded semantic search. On the background of development of HyperGraph - Graph (HG(2)) data structure, an RDF document if integrated to it, maintains the original semantics and exposes some critical semantic and object mapping lift as well which could further be exploited for semantic search and information transitional problems. The focus of the paper is to present the mapping constructs between RDF elements and HyperGraph - Graph (HG(2)) elements.

## I. INTRODUCTION

As suggested by Tim Berners-Lee et al [1], the semantic web is considered to be different from web 2.0 as the target au-diences of the former are the machines instead of humans. As a result, semantic web or web 3.0 is thought to be a huge network of information that can be interpreted by machines. Looking back to the initial periods of web 2.0 until XML was pro-posed, it was evident that problems of data manipulation were primarily due to the non-existence of common data framework. Similarly, current web 3.0 infrastructure which is characterized by co-existence of multiple information framework, expects a common information framework for seamless exchange of information across the different information framework types. Resource Description Framework (RDF) and Topic Map are currently, the mostly used semantic web information standards. These two frameworks, as patronized by two different stan-dardization giants (W3C and ISO respectively), have grown up differently to limit the scope of accessibility and readability of semantic web information as a whole.

On the basis of the consideration of the fact that the co existence of multiple information framework limits the potential of Semantic Web, current paper is driven by the idea presented in [7] which establishes that information integration of different semantic metamodel into a single data structure is a better alternative than designing multiple cross platform information transition schemes. The monolithic data structure for information integration has been proposed in [7] as Hypergraph-Graph (HG(2)) and its design criteria and theoretic analysis have been reported in [8, 9].

On the background of theoretical development of Hypergraph - Graph (HG(2)) data structure, the bottleneck in designing a platform for semantic information integration appears as the absence of concrete mapping scheme from different semantic framework elements into HG(2) constructs. Said this, the motivation of the current paper could be summarized as to present and analyze the mapping scheme of RDF and RDF Schema elements into HG(2) framework. The mapping investigated involving RDF element is complete but the RDF Schema element mapping have scopes of further inclusions.

The present paper has been structured as follows. Section 2 reviews the fundamental concepts of RDF and RDF Schema in separate subsections. The design and theoretical considerations of Hypergraph - Graph (HG(2)) data structure are presented in Section 3, which is followed by introduction of mapping schemes of RDF and RDF Schema/Vocabulary elements to HG(2)) components in Section 4. The current study ends with logical discussions on future research scopes and conclusion in Section 5.

## II. OVERVIEW OF RDF AND RDF SCHEMA

The Resource Description Framework (RDF) is a simple metamodel that can define, construct and exchange information on the semantic web. Tim Berners and Lee were the first to officially propose the concepts of RDF and currently it is being evolved as a W3C standard [10]. RDF Schema, however, can best be described as the fixed set of rules that defines compositional and structural component of RDF. Following are the two subsections which present the overview of RDF and RDF Schema.

### A. Resource Description Framework (RDF)

RDF is simple, platform neutral information metamodel which is composed of statements from an unordered set [11]. Each statement can be thought of as a triple. The triple designates a relation between a subject and an object through a predicate. The subject of each statement is considered to be a resource which can represent any-thing identifiable in this universe. Each of the Resources can be of two types, either it has a single global URI identifier or it is a 'blank' resource. A Blank Resource exists with unique identity but there exists no identifier for the same. A predicate describes the relationship between the subject and the object and it is also considered to be a resource. As a result, a predicate is also allowed to be blank. The object of each statement is thought to be either a resource or a literal. Literals are described as strings and it might have a language tag associated with it. Hence a literal can be interpreted either as a simple string when it is represented as a String object or as an XML fragment when it is associated with a language tag. A literal is not allowed to be represented the subject of a statement.

As per W3C study, RDF could best be utilized as a modeling tool for representing the semantics of data or information that are implemented in web resources. RDF can also be described through a simple but strictly-defined textual format for graph data structure where each of the nodes may represent Subject or Object and edges represent Predicates. Such graphs that describes triple structure of RDF are denoted as RDF Graph. There exists a separate nomenclature as N-Triple which is a specific format for representing triple structure of RDF. The following are the three N-Triples (as shown in W3C site) which describe of three different RDF statements.

TABLE I. N-TRIPLE REPRESENTATION OF RDF

| N-Triples | |
|---|---|
| 1 | Subject:<http://www.w3.org/2001/sw/RDFCore/ntriples/> Predicate:< http://purl.org/dc/elements/1.1/creator> Object: "Dave Beckett" |
| 2 | Subject:<http://www.w3.org/2001/sw/RDFCore/ntriples/> Predicate: <http://purl.org/dc/elements/1.1/creator> Object: "Art Barstow" |
| 3 | Subject:<http://www.w3.org/2001/sw/RDFCore/ntriples/> Predicate: <http://purl.org/dc/elements/1.1/publisher> Object: <http://www.w3.org/> |

## B. RDF Schema

There exist many literatures exploring RDF Schema in details. The previous studies suggest that within the RDF metamodel itself, there exists no means for describing the properties associated with the resources. Moreover, there is no mechanisms for designating the inter-relationships between the properties and the resources. Hence RDF Schema provides a specification which just represents the set of rules to describe properties and to describe the relationship between resource and property. There are two fundamental concepts RDF Schema ( RDFS in short): Property and Type. Type is considered as a resource used as the predicate when describing the type or class of another resource. On the other hand, Property is conceived as the type of all resources that can be used as predicates. Hence RDFS is considered to be the language for RDF vocabulary description.

RDFS incorporates the notion of typing through introduction of the concept of Class. The type of every resource must be an instance of Class. Class is itself of type Class, hence RDFS is identified as an unstratified structure. RDFS suggests that every resource must be the instance of at least one Class, and hence it identifies Resource as the class of all resources and the class Literal as the class of all literals.

The RDFS is composed of some basic classes and properties, and can be extended by others according to the current domain. Classes maintain hierarchical structure, and only the members of certain classes can use properties. The root of the class hierarchy is rdfs:Resource, and rdfs:Class is subclass of rdfs:Resource

Properties are defined by the rdf:Property class and can be considered to be attributes, which describe resources by assigning values to them. RDFS also defines four specific properties (rdfs:subClassOf, rdf:type, rdfs:range, rdfs:domain) that have certain constraints. Additional predefined properties such as rdfs:seeAlso and rdfs:comment are used to provide a human readable description of a resource.

## III. HYPERGRAPH - GRAPH (HG(2)) DATA STRUCTURE

Hypergraph - Graph data structure denoted as HG(2) is conceptualized as a model to represent a complex problem space based on certain criteria. The criteria could be formalized as follows -
The problem space ($PS$) must logically be divided into two levels with different complexities, one ($PSG$) with relatively lesser complexities, better orderdness and bounded by for-malized set of rules, and another ($PSH$) which could be characterized by greater complexities and absence of ordered rule sets.
The some or all interrelationship between objects of $PSH$ must be dictated by the objects of $PSG$ and even the behavior of some or all objects of $PSH$ must be defined by the objects of $PSG$. Here the term "object" is being used informally and must not necessarily indicate any Object Oriented paradigm. As the complex real life combinatorial structures are not rare at all, proposed HG(2) has an intrinsic objective to represent $PSH$ with Hypergraphs and $PSG$ with Graphs. The inter dependencies of $PSH$ and $PSG$ form the basis of evolution of the behavior of HG(2) as a whole.

The theories behind the Hypergraph Data Structure are presented in [6] and due to space constraint, it is omitted in the current discussion. On this background, following subsection presents the theories of Hypergraph - Graph (HG(2)) data structure directly.

### A. Introducing HG(2)

A Hypergraph-graph data structure HG(2) is a triple denoted as $HG(2) = (H, G, C)$ where $H$ is a Hypergraph, $G$ is a graph and $C$ is a set of *connectors*.

$H$ is a Hypergraph defined as $H = (V^h, E^h)$, where $V^h = v_1^h, v_2^h, \cdots, v_n^h$, $n = 2, 3, \cdots$ and $E^h = E_1^h, E_2^h, \cdots, E_m^h$, $m = 2, 3, \cdots$ where each $E_i \subseteq V$
$G$ is a Graph defined as $G = (V^g, E^g)$, where $V^g = v_1^g, v_2^g \cdots v_p^g$, $p = 2, 3, \cdots$; and $E^g = e_1^g, e_2^g \cdots e_q^g$, $q = 2, 3, \cdots$ where each $e_i$ could be expressed in the form of $e_{xv}$ which connects $v_y^g$ from $v_x^g$.

$C$ is a set of *connectors*, which could be conceptualized as a set of all the dependencies between $PSH$ and $PSG$ (as described earlier) which are characterized by $H$ and $G$ respectively.

here we define two types of connectors:
(a) $c^v_{xy}$ which connects a node in the Graph $v_y^g$ from a

node in the Hypergraph $v_x^h$. It is to note that the behavioral dependency of an object of $PSH$ on an object of $PSG$ gets realized through $c_{xy}^v$; and

(b) $c_{xy}^e$ which connects a node in the Graph $v_y^g$ from a Hyperedge $E_x^h$. Here $c_{xy}^e$ realizes the dependencies of collective behavior (bound with a specific relation) of the objects of $PSH$ on the objects of $PSG$.

The set of all $c_{xy}^v$ s is noted as $C^v$ while $C^e$ represents all $c_{xy}^e$ s. Hence on the basis of ongoing discussion, it could be concluded that $C = (C^v, C^e)$. For the rest of the paper, it is assumes for simplicity that the dependency flows from $PSH$ to $PSG$. That is the Hypergraph layer is dependent on the Graph layer. No dependency flows through a connector backward from the Graph layer to the Hypergraph layer. Above discussion can be illustrated with the example as shown in Fig 1. In this figure, an HG(2) is shown to have a Hypergraph $H$ and a Graph $G$. The $H$ and the $G$ are connected with *connectors* $C$. The Hypergraph $H$ is composed of ($V^h, E^h$) where $V^h$ consists nodes 1, 2, 3, 4, 5, 6 & 7 and $E^h$ consists $E_1, E_2, E_3$ & $E_4$. Further, the Graph $G$ is composed of nodes $a, b, c, d, e$ & $f$. For simplicity the edges of the $G$ are not highlighted.

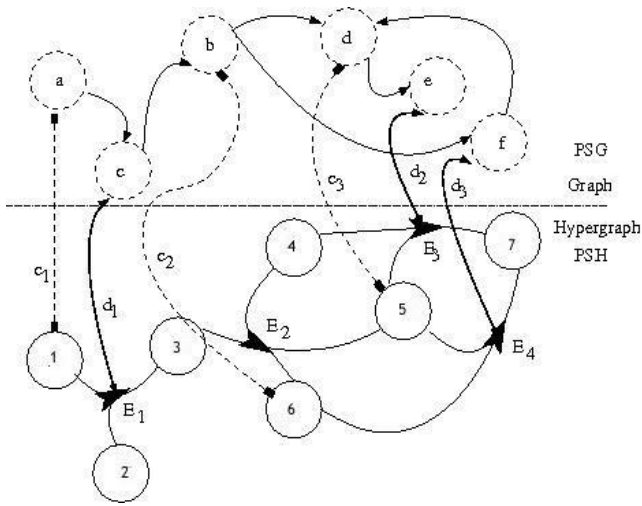

Fig. 1. Illustration of an HG(2) data structure

It is to be noted that $E_1$ is composed of nodes 1, 2, 3, $E_2$ is composed of nodes 3, 4, 5, 6, $E_3$ is composed of nodes 4, 5, 7 and finally $E_4$ is composed of nodes 5, 6 and 7. Following is the identification of head and tail nodes for each Hyperedge:
$H(E_1) = 1, 2$ and $T(E_1) = 3$
$H(E_2) = 3, 4$ and $T(E_2) = 5, 6$
$H(E_3) = 4, 5$ and $T(E_3) = 7$
$H(E_4) = 5, 6$ and $T(E_4) = 7$

The Fig 1 further illustrates that there are three connectors (marked with dashed square ended bi-directional arrow) that connect Hyperedge nodes and Graph nodes. They are $c_{1a}^v = c_1$, $c_{6b}^v = c_2$, $c_{2d}^v = c_3$. Moreover there are three connectors (marked with bold bidirectional arrow) that connect a Hyperedge with a Graph node. They are identified as $c_{1c}^e = d_1$, $c_{3e}^e = d_2$ and $c_{4f}^e = d_3$. Hence formally $C$ which is a pair ($C^v, C^e$) holds the following:
$C^v = c_1, c_2$ and $c_3$; and $C^e = d_1, d_2$ and $d_3$.

Though the diagram shows the individual connectors with bidirectional arrow, the earlier assumption still holds that the direction of the dependency is from Hypergraph to Graph layer. The bidirectional arrows have been given for easy diagrammatic identification of connectors out of many different edges.

However, Fig 1 shows the entire problem space to be logically divided (by a dashed horizontal divider) into $PSG$ and $PSH$ which are represented by the Graph and Hypergraph respectively.

On the background of the discussion so far, following section presents a clear approach of mapping RDF elements to HG(2) elements.

IV. RDF ELEMENTS TO HG(2) ELEMENTS MAPPING

As HG(2) has two distinct layers of complexities realized by Graph and Hypergraph, the RDF NTriple elements that is, Subject, Object, Predicate, Literals and Blank Nodes, all are needed to be mapped to Hypergraph components due to their intrinsic complexities and absence of structured associations. Just opposite to this, the structured knowledge level of RDF Schema could be represented by Graph components. Hence the broad design criteria could be fixed as follows-
(a) RDF Schema maps to Graph Data Structure
(b) RDF Graphs map to Hypergraph Data Structure.

The criteria just stated separates RDF primitives and RDF Schema logically. The two logical segments of HG(2) then needs to be bind with the logical entities known as Connectors. The connectors of HG(2) actually exist any represent the assertions of the fact whether a hypergraph component can have any logical association with the graph component. If there exists any logical association as per the RDF specification between RDF primitives and RDF Schema, a valid connector will exist in between.

On the light of this broad criteria, following subsections investigates the detailed granular level mapping.

*A. Mapping of RDF primitives*

RDF primitives as per the RDF specification could be identifies as Statements, Subject, Object and Predicate. A Statement is composed of a Subject and an Object related with a Predicate. Hence a Statement could be mapped to a Hyperedge $E_i^h$ and the set of multiple number of such Statements could be formalized as $E^h$. By definition of a Hyperedge with cardinality $n$ is composed of $n$ number of Hypernodes. As the cardinality of an RDF Statement is always three, there will be always three hypernodes associated with a Statement. The other three RDF primitives as Subject, Object and Predicate could then individually be mapped as Hypernodes $V^h$ s. For the sake of simplicity and general ordering, a Predicate is considered to be Head Node of any Hyperedge and a Subject and an Object are considered as two Tail Nodes of a Hyperedge.

W3C specification on RDF dictates that a Subject or an Object or a Predicate could always be realized with URI Ref-erences. Hence any of the $V^h$'s could potentially be identified with URI Reference. On the contrary, A Literal can only be represented as an Object, neither as a Subject, nor as a Predicate. Hence a Head Node could never map a Literal but a Tail Node can (as an object).

Blank Nodes in RDF can represent an Object of a Statement which works as a Subject of another Statement. Hence within a Hypergraph if a Tail Node of a hyperedge works only as Tail Node of another hyperedge, it could map a Blank Node

Recursively, a Plain Literal could be composed of a Lexical Form and an optional language tag, and a Typed Literal could be composed of Lexical Form and Datatype URI. As a result, a Head Node must not be mapped with Lexical Form and Language Tag, but a Tail Node may have a Lexical Form or a Language Tag.

The above discussion could be formalized in Table 2 as described below.

TABLE II. MAPPING OF RDF PRIMITIVES TO HYPERGRAPH ELEMENTS

| RDF Primitives | Hypergraph Elements |
|---|---|
| Statement | Hyperedge |
| URI Reference | both Head and Tail Nodes of a hyperedge |
| Blank Node | Only Tail Nodes of a Hyperedge working only as Tail Node of another Hyperedge |
| Lexical Form | Only Tail Nodes of a Hyperedge |
| Data Type URI | Only Tail Nodes of a Hyperedge |
| Language Tag | Only Tail Nodes of a Hyperedge |

### B. Mapping of RDF Schema Constructs

Figure 2 shows the hierarchical structure of the RDF data model. If a class is inherited from another, then there is an rdfs:subClassOf arc from the node which represents the daughter class to the node representing the parent. However, if a Resource is an instance of a Class, then there exists an arc rdf:type from the resource to the node representing the class. However, rdfs:subClassOf can exist both as a specific property and a primitive construct which is designated as an arrow labeled with s. Similarly, rdf:type can exist both as specific instance of property and as primitive construct which is designated as an arrow labeled with t. Similarly Figure

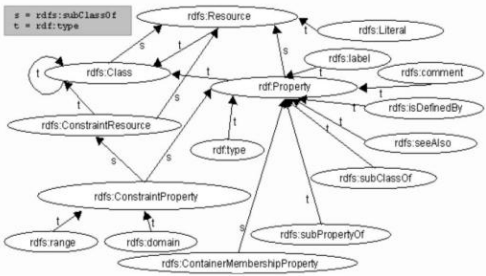

Fig. 2. Class Hierarchy of RDF Schema

3 describes the RDFS meta structure. Here rdfs:range and rdfs:domain constrain the relationship between RDF classes and properties. Hence rdfs:range and rdfs:domain can both be implemented as as implicit constructs and explicit properties as well. The formal description of the RDF Schema as shown in

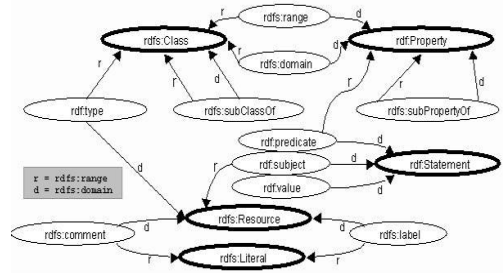

Fig. 3. Constraints in RDF Schema

Fig 2 and 3 generates enough motivation for representing RDF Schema with a Graph within a Hypergraph. Any Graph node $v_i^g$ maps rdfs classes and there could be four types of graph edges $e_i^{g_s}$, $e_i^{g_t}$, $e_i^{g_d}$ and $e_i^{g_f}$ corresponding rdfs:subClassOf, rdf:type, rdfs:domain and rdfs:range respectively.

On this background, current paragraph relates the RDF Schema constructs and RDF Primitives in the form of Con-nectors of HG(2). As described in the previous section, there exists two types of Connectors: Node to Node ($C_{xy}^v$) and Edge to Node ($C_{xy}^e$). While integrating the RDF to HG(2), Edge to Node Connectors will map the the logical assertiveness of RDF Statement being rdf:statement. Similarly, the logical association of some hypernodes and graph nodes will be mapped via Node to Node Connectors. For example Node to Node connector will connect Tail Nodes (of an hyperedge or RDF Statement) to rdf:object, Head Nodes to rdf:predicate, Data type URI to rdf:datatype etc.

The significance of integrating RDF document in HG(2) with the mapping techniques as discussed can be attributed to the fact that the semantic expression of RDF and structural completeness of RDF Schema are tied within a monolithic structure. This results in better syntax to semantic mapping and object level recognition of RDF. Integration of RDF to HG(2) opens up bright avenues of mapping other semantic metamodels to HG(2) as well. Thus HG(2) could act as a integration platform for different semantic web metamodels and could potentially facilitate better information portability and broader search. The walled (RDF, Topic Map, OWL etc) garden of Semantic Web could effectively be bridged with further research on HG(2) and metamodel integration/mapping mechanism.

## V. CONCLUSION

The present research has introduced a mapping schemes of RDF and RDF Schema elements into HG(2) to integrate RDF documents with a common information platform. The mapping procedure presented formally preserves the semantic of RDF vocabulary and exposes further semantic lift in object domain which could be exploited with Path Traversal algorithms on HG(2).

Within the current domain of research, the future scope of RDF integration in HG(2) has aptly been identified to focus on applying Path Traversal and Minimum Spanning Tree Algorithm on HG(2) for semantic searching on Web.